# A classical approach to the electron g-factor

Jaromír Chalupský*

*Institute of Physics, Academy of Sciences of the Czech Republic, Na Slovance 2, 182 21 Prague, Czech Republic*

*E-mail: chal@fzu.cz

**ABSTRACT**

According to a prevailing opinion, the electron g-factor $g_e$ = 2 is exclusively a quantum feature. Here we demonstrate it could be explained classically only in relativistic terms. The electron is treated as an extended, continuous, but rigid Gaussian body (RGB) spinning at finite angular frequency. In contrast to expectations, the mechanical energy and spin angular momentum of the particle are not diverging but standard values are reproduced. The g-factor value $g_e$ = 2 immediately follows from the ratio of non-relativistic and relativistic angular momenta which can be both attributed to a spinning electron of known rest mass. A detailed analysis of the inertia tensor and limit, torque-free precession reveals a multiplication factor of -2 between the external and internal precession angular frequency which might resemble the spin-1/2 appearance of the particle. Furthermore, the theory of Liénard and Wiechert is used to derive a static electromagnetic field. A continuous form of Gaussian charge density ensures an absence of infinities in electromagnetic energy and angular momentum. Introducing the associated electromagnetic angular momentum as a small correction to the mechanical spin angular momentum, we obtain a modified g-factor $g_e$* = 2.0021 which is close to the measured value $g_e$ = 2.0023.

## 1. Introduction

The discovery of half-integer electron spin by Uhlenbeck and Goudsmit [1] represented a real revolution in early quantum mechanics. The theory immediately led to a resolution of "duplexity" phenomena in hydrogen-like spectra being a hot topic at that time. After confirmation by the experiments of Stern and Gerlach [2] and Phipps [3], the spin was accepted as the fourth quantum number, making the quantum picture of an atom almost complete. However, this solution brought other difficulties in understanding underlying physical principles, from which the particle spin could emerge. The electron was believed to be point-like which prevented classical approaches to its spin angular momentum (SAM) and kinematics in external fields. Quantum mechanics was, by far, more successful in the description of electron kinematics in the electromagnetic field. The first attempt to involve the spin into the Schrödinger equation was carried out by Pauli in 1927 [4]. This modification is known as the non-relativistic Pauli equation operating on spinor wave-functions. Nevertheless, Dirac pointed out that some inconsistencies could follow from the incompatibility of Pauli's equation with special relativity. With some mistrust to the Klein-Gordon equation [5,6], Dirac derived his relativistic wave-equation [7] from which the g-factor $g_e$ = 2 of an electron naturally emerges [8]. This value can be regarded as the first approach since, according to quantum electrodynamics, there are higher-order corrections to the g-factor, usually termed as the anomalous magnetic dipole moment, making the g-value greater than two. These corrections emerge from the fact that the electron is not isolated from the surrounding environment but may interact with virtual photons of QED vacuum and with its own electromagnetic field. The QED-predicted value of the g-factor ($g_e$ = 2.0023193043768) [9,10] is in excellent agreement with latest measurements ($g_e$ = 2.0023193043617) [11,12] which makes QED the most precisely tested theory in the history of physics. The success of quantum electrodynamics led to a widespread belief that the electron spin and g-factor is purely of quantum nature without any classical analogue. Despite this strict conclusion, classical and other "unconventional" approaches to the electron spin and the underlying phenomena are still ongoing; see, for example, works of Cohen [13], Barut [14], Ohanian [15], MacGregor [16], Ghosh [17], Božić [18] and Czachor [19]. Evidently, an unflagging interest in alternative explanations of the electron spin indicates that our understanding of this topic is still not complete.

In this paper we demonstrate a classical relativistic model of an electron which is capable of predicting the g-factor to a very good accuracy in a non-quantum manner. The model assumes that the rest mass and charge of an electron are continuously but rigidly distributed in space, both following the same Gaussian probability density. At this stage we focus only to a static (non-moving) particle which rotates around a fixed axis at constant angular frequency. There are, however, three intuitive assumptions we have to accept without rigorous scientific justification, namely, the choice of continuous Gaussian mass and charge density, the rigidity (compactness) of the spinning particle, and the existence of superluminal and luminal velocities at radii greater than or equal to a certain critical radius.

The choice of the Gaussian body is mathematically (statistically) and empirically motivated[1]. It cannot be rigorously justified unless we understand the meaning of the probability density giving the particle its existence and shape[2]. The rigidity of the particle intuitively emerges from its continuous probabilistic nature and steady rotational motion in a closed loop. In fact, if we make a continuous, smooth, and rotationally symmetric object (mass/charge density) to spin around the symmetry axis, we cannot even recognize if it rotates or not since no distinguishable and traceable markers exist. Such an object appears as static and rigid but its spin angular momentum and magnetic dipole moment is non-zero. The presented model is an example of Born rigidity [20] defining a rigid body in terms of special relativity. This concept was subject to criticism, extensions, and modifications from both the kinematical and dynamical standpoint. Kinematical approaches were mostly concerned with Lorentz transformation [21-24] and rigid motions of rotating non-inertial reference frames. It was confirmed by Herglotz [21] that a rotating rigid body may exist; however, Cantoni [22] showed that the rigidity cannot be maintained during the phase of angular acceleration from rest to steady rotation. Ehrenfest's [23] objection to Born rigidity was that the circumference of a rotating relativistic body must undergo Lorentz contraction whereas the radius stays intact. Nevertheless, as it was pointed out by Grøn [24], the weak point of Ehrenfest's paradox resides in the phase of acceleration which cannot be described in terms of special relativity.

---

[1] For example, in quantum mechanics the particles are very often described as Gaussian wave-packets exhibiting the lowest possible Heisenberg uncertainty.
[2] The existence of the particle must be invariant with its velocity. Therefore, the integral of the probability density function over the entire space is unity even at relativistic speeds. The shape of the particle undergoes the Lorentz contraction while maintaining the unitary probability of existence.

The Born-rigid body was disputed also from the dynamical point of view [24-27] since any extended body approach encounters the well-known issue of centrifugal and stress forces leading to deformation. It was believed that the rigid body violates fundamental relativistic principles since the speed of propagation of cohesive forces (e.g., speed of sound), keeping the body rigidly compact, would have to be infinite [27]. However, this counterargument is based on the concept of discrete particle ensembles kept together by particle-particle interactions. In the presented continuous picture, no independent and mutually interacting substances can be identified within an electron; on the contrary, the electron itself is to be regarded as a fundamental substance of matter. Resolution of this question is definitely not a simple task; therefore, as stated at the beginning, let us assume that an extended electron in steady motion is rigidly compact and concede the fact that an abrupt nonstationary change to the state of motion most likely disturbs the rigid appearance of the particle.

Real examples of Born-rigid motions exist. It follows from the Liénard-Wiechert theory [28] that the electric and magnetic field of a charge in uniform (steady) motion rigidly follows the trajectory of the particle within a sphere in which the information about the current state of the particle motion is available. Accelerated motions, collisions, and other abrupt changes in the particle's state of motion affect the rigid appearance of the electromagnetic field and may lead to a creation of secondary particles, e.g., photons. In the frame of the RGB model we will show that the electron in steady rotation around its symmetry axis creates a static electromagnetic field but does not radiate. Of course, it could be objected that, for example, the electron current density in a bending magnet can also be steady but synchrotron radiation is detectable. The fundamental difference, however, resides in the fact that the current density in the bending magnet is not continuous but discrete, consisting of individual orbiting electrons radiating due to an accelerated motion, i.e., due to the change of the state of motion. On the contrary, in the rotating electron the current density is continuous and steady which prevents radiation losses.

The last intuitive assumption in question is the existence of superluminal and luminal velocities at radii greater than or equal to a certain critical radius. Since the rigid Gaussian body is infinitely extended, it has to deal with consequences of this issue even at very small angular velocities. One should be concerned with what happens to the mass (energy) density in the space outside and at the critical radius where the local velocity is greater than or equal to the speed of light, respectively. Discrete ensembles of point-particles are automatically excluded from this question since the energy of particles localized at the critical radius diverges. On the contrary, a continuous mass density rotating around its symmetry axis may give a finite, albeit complex, value of the angular momentum and energy, as we show later. Furthermore, we still have to answer the problem of superluminosity. It was stated above that the rotation of a rotationally symmetric and continuously distributed rigid body cannot be distinguished from the static case unless the spin angular momentum or magnetic dipole moment is measured. Since the particle appears static, no information or energy is transferred to an observer unless the steady state of the particle rotation is disturbed, e.g., the orientation of the spin is flipped. In this case a photon can be emitted and information about the change will propagate through the space at the speed of light. Superluminal velocities are not strictly excluded by the special theory of relativity unless some information or energy is transmitted.

Accepting the above stated assumptions, we may derive the properties of a spinning rigid Gaussian body. First, we calculate the energy and the spin angular momentum of the rotating electron in the relativistic and non-relativistic regime. From here the g-factor ($g_e$ = 2) will immediately emerge. Next, we construct the tensor of inertia and discuss how the limit precession (infinitesimally small precession angle) might correspond to the spin-1/2 appearance of an electron. Then we derive the electric and magnetic field of the rotating charged electron from which we obtain finite values of electromagnetic energy and angular momentum. By considering the electromagnetic angular momentum as a small correction to the mechanical spin angular momentum, we obtain the g-factor to an improved accuracy ($g_e^*$ = 2.0021).

## 2. Mechanical angular momentum and energy

Let us first define the mass density of a rigid Gaussian body, which is by definition spherically symmetric:

$$\rho_m(\mathbf{r}) = \frac{m_0}{\left(\sqrt{\pi}R\right)^3} \exp\left(-\frac{|\mathbf{r}|^2}{R^2}\right), \tag{1}$$

where $\mathbf{r} = (x,y,z)$ is the position vector, $m_0$ = 9.1094·10$^{-31}$ kg is the electron rest mass, and $R$ is the radius of the density function at 1/e of maximum. The mass density function is normalized in such a way that it's integral in the whole space is equal to the rest mass $m_0$ of the electron.

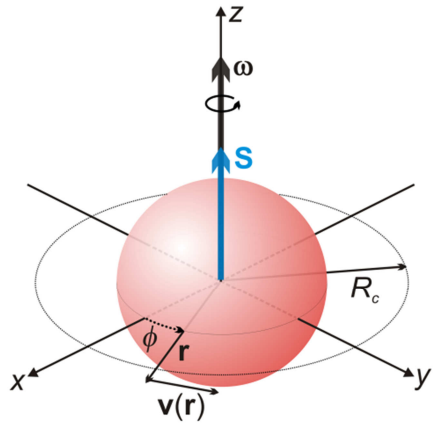

Fig. 1 – Orientation of the spinning body in Cartesian coordinates. Vectors **ω**, **S**, **r**, and **v(r)** denote the angular velocity, spin angular momentum, local position, and local velocity, respectively. Radius $R_c$ stands for the critical radius.

As shown in Fig. 1, at each point of the space **r** we can define a vector of local velocity **v(r)** and scalar value of Lorentz factor $\gamma(\mathbf{r})$ which, according to the requirement of rigidity and steady rotation, read:

$$\mathbf{v}(\mathbf{r}) = c\boldsymbol{\beta}(\mathbf{r}) = \boldsymbol{\omega} \times \mathbf{r}, \tag{2a}$$

$$\gamma(\mathbf{r}) = \left(1 - |\boldsymbol{\beta}(\mathbf{r})|^2\right)^{-1/2}. \tag{2b}$$

Here $c$ = 299792458 m/s is the speed of light in vacuum and **ω** is the vector of angular velocity defining the axis of rotation. The condition of rigid rotation leads to a conclusion that there exists a critical cylindrical surface of radius $R_c = c/\omega$ at which the local velocity reaches the speed of light. Inside the cylinder the velocity is subluminal and the Lorentz factor is real whereas outside the critical cylinder superluminal speeds result in imaginary Lorentz factors. Therefore, the energy and spin angular momentum of the particle will obtain an additional imaginary part. Even though this approach may be mathematically correct, contemporary physics does not recognize imaginary energies or momenta; therefore, we will disregard the imaginary parts.

The spin angular momentum and energy of the rotating electron can be expressed in integral forms:

$$\mathbf{S} = \int_{R^3} \mathbf{r} \times d\mathbf{p} = \int_{R^3} \mathbf{r} \times \mathbf{v}(\mathbf{r}) \gamma(\mathbf{r}) \rho_m(\mathbf{r}) d^3 r, \tag{3a}$$

$$E = \int_{R^3} c^2 dm = \int_{R^3} c^2 \gamma(\mathbf{r}) \rho_m(\mathbf{r}) d^3 r, \tag{3b}$$

where we integrate over the entire space. Due to the cylindrical symmetry, the integrals can be more conveniently solved in cylindrical coordinates where, without loss of generality, one may orient the axis of rotation parallel to the z-axis, as depicted in Fig. 1. In this geometry the spin angular momentum **S** will retain only the z-component $S_z$. If we express the positional vector as $\mathbf{r} := (r\cos(\phi), r\sin(\phi), z)$, transform the integrals in (3a) and (3b) to cylindrical coordinates, and insert from definitions (1), (2a), and (2b), we obtain the following analytical complex solutions:

$$S_z(B) = m_0 cR \left\{ \frac{B^2 + 2}{B^2} \left[ F\left(\frac{1}{B}\right) - i\frac{\sqrt{\pi}}{2} \exp\left(-\frac{1}{B^2}\right) \right] - \frac{1}{B} \right\}, \tag{4a}$$

$$E(B) = m_0 c^2 \left\{ \frac{2}{B} F\left(\frac{1}{B}\right) - i\frac{\sqrt{\pi}}{B} \exp\left(-\frac{1}{B^2}\right) \right\}, \tag{4b}$$

where $B = \omega R/c = kR$ is the velocity parameter, i.e., local velocity at $1/e$ radius $R$, $k$ is the wavenumber corresponding to the angular frequency $\omega$, $i$ is the imaginary unit ($i^2 = -1$), and F(x) is the so called Dawson's function [29], being real for real arguments. The Dawson function is related to the imaginary error function erfi(x) and error function erf(x) through the following expressions:

$$F(x) = \frac{\sqrt{\pi}}{2} e^{-x^2} \operatorname{erfi}(x) = -i\frac{\sqrt{\pi}}{2} e^{-x^2} \operatorname{erf}(ix). \tag{5}$$

Dependencies of the spin angular momentum and particle energy on the velocity parameter $B = kR = \omega R/c$ are depicted in Figs. 2(a) and 2(b). Evidently, both the spin angular momentum and energy are finite and do not diverge as $B$ tends to infinity. As one could expect, the spin angular momentum is equal to zero for $B_0 = \omega R/c = 0$ since, in this particular case, the particle is either point-like ($R = 0$) or not in rotation ($\omega = 0$). Furthermore, the real parts of spin angular momentum and energy are odd and even functions of $B$, respectively. Both quantities reach a local maximum at certain $B$ and drop to zero as $B$ tends to plus/minus infinity. As naturally follows from Eq. (1), the energy equals to the rest energy ($E = m_0 c^2$) for $B_0 = 0$, i.e., for a non-rotating and/or point-like particle. However, the equation $\operatorname{Re}\{E(B)\} = m_0 c^2$ has two more solutions corresponding to an extended particle in clockwise or counterclockwise rotation, as shown in Fig. 2(b). By solving this equation numerically, we obtain $B_\pm = \pm 1.082088449$ for which the spin angular momentum is non-zero and energy equal to the rest energy. Since it holds true that $F(1/B_\pm)/B_\pm = 0.5$, we can express the real parts of these quantities as[3]:

$$S_z(B_\pm) = S_z^\pm = m_0 cR \frac{B_\pm}{2}, \tag{6a}$$

$$E(B_\pm) = E_0 = m_0 c^2. \tag{6b}$$

In summary, it follows from this analysis that a rigid Gaussian particle, which is set into a rotation, has finite energy being under certain conditions equal to the rest energy. In addition to this, the particle gains a non-zero spin angular momentum. Therefore, both quantities, i.e., the rest energy and spin angular momentum, seem to obtain their values due to the fact that the particle is spinning.

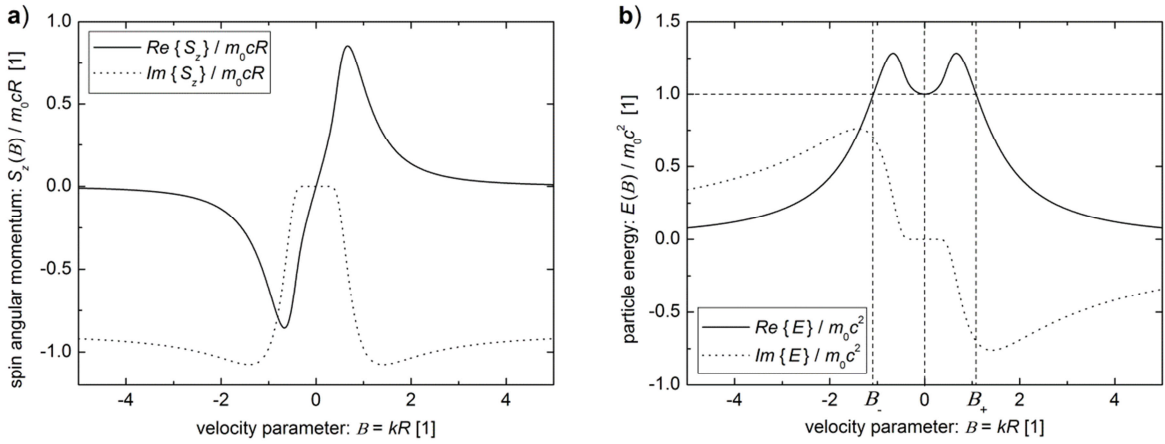

Fig. 2 – (a) Real and imaginary part of spin angular momentum as a function of velocity parameter $B = kR$. (b) Real and imaginary part of particle energy as a function of velocity parameter. Only real parts will be considered in the following calculations.

For the purpose of further analysis, let us explore the asymptotic behavior of the spin angular momentum in the non-relativistic limit, i.e., for $kr \ll 1$. In order to do so, we have to neglect the Lorentz factor (2b) in Eq. (3a) which will thereby have the form:

$$\mathbf{S}^{nR} = \int_{R^3} \mathbf{r} \times \mathbf{v}(\mathbf{r}) \rho_m(\mathbf{r}) d^3 r. \tag{7}$$

Again, by solving this integral in cylindrical coordinates for the mass density (1) and velocity field (2a), we obtain the z-component of the non-relativistic spin angular momentum:

$$S_z^{nR}(B) = m_0 cRB. \tag{8}$$

Here we recall the velocity parameter $B = \omega R/c = kR$. Evidently, the spin angular momentum in the non-relativistic limit $S_z^{nR}(B)$ is a linear function of $B$ being tangent to its relativistic analogue $S_z(B)$ at $B = 0$.

---

[3] From this moment on, we disregard the imaginary parts of the spin angular momentum, energy, and other related quantities.

## 3. Derivation of the g-factor ($g_e = 2$)

In order to express the g-factor of an electron, we start with a general definition of the magnetic dipole moment (c.f. [28]):

$$\boldsymbol{\mu} = \frac{1}{2}\int_{R^3} \mathbf{r} \times \mathbf{j}_e(\mathbf{r}) d^3 r = \frac{1}{2}\int_{R^3} \mathbf{r} \times \mathbf{v}(\mathbf{r})\rho_e(\mathbf{r}) d^3 r. \tag{9}$$

Here **v(r)** is the vector field of velocity, as defined in (2a), and $\rho_e(\mathbf{r})$ and **j**$_e$(**r**) are the charge and current density of the spinning rigid Gaussian body:

$$\rho_e(\mathbf{r}) = \frac{q}{\left(\sqrt{\pi}R\right)^3} \exp\left(-\frac{|\mathbf{r}|^2}{R^2}\right) = \frac{q}{m_0}\rho_m(\mathbf{r}), \tag{10a}$$

$$\mathbf{j}_e(\mathbf{r}) = \rho_e(\mathbf{r})\mathbf{v}(\mathbf{r}) = \rho_e(\mathbf{r})\boldsymbol{\omega} \times \mathbf{r}, \tag{10b}$$

where $q = -1e = -1.6022 \cdot 10^{-19}$ C is the elementary charge of an electron. The charge density is assumed to be of the same shape and 1/e radius as the mass density in Eq. (1).

The magnetic dipole moment in Eq. (9) can be expressed in terms of the non-relativistic spin angular momentum (7) since the integrals, involved in both equations, are equivalent, apart from some multiplication factor. In fact, this is a standard method how to connect the magnetic dipole moment with the non-relativistic spin angular momentum. Therefore, the z-component of the magnetic dipole moment, as a function of the velocity parameter $B = kR$, reads:

$$\mu_z(B) = -\frac{\mu_B}{\hbar} S_z^{nR}(B), \tag{11}$$

where $\mu_B = e\hbar/2m_0 = 9.2740 \cdot 10^{-24}$ J/T is the so called Bohr magneton and $\hbar = 1.0546 \cdot 10^{-34}$ Js is the reduced Planck constant.

In the previous section we calculated that an electron spinning with the velocity parameter $B_\pm = \pm 1.0821$ will possess the spin angular momentum and rest energy given by Eqs. (6a) and (6b). Clearly, the velocity parameter $B_\pm$ significantly exceeds the non-relativistic regime and, therefore, the correct value of SAM is represented solely by Eq. (6a). It immediately follows from Eqs. (6a) and (8) that $S_z^{nR}(B_\pm) = 2S_z(B_\pm)$. This makes it possible to re-evaluate the magnetic dipole moment (11) in terms of the relativistic SAM as:

$$\mu_z(B_\pm) = -2\frac{\mu_B}{\hbar} S_z(B_\pm) \Rightarrow g_e = \frac{S_z^{nR}(B_\pm)}{S_z(B_\pm)} = 2, \tag{12}$$

whereby, the origin of the electron g-factor $g_e = 2$ is justified classically. In essence, the emergence of the g-factor is due to the fact that the charge is a relativistic invariant whereas the mass is not. Hence the link between the magnetic dipole moment and spin angular momentum cannot be expressed in simple terms of Eq. (11). This equation is only mathematically correct; nevertheless, its physical interpretation is inconsistent unless relativistic effects on the spin angular momentum are involved.

## 4. Tensor of inertia and limit precession

Let us develop the theory of the relativistic spin angular momentum a little further and explore Eq. (3a) in more detail. For this purpose we return to Cartesian coordinates, i.e., **r** = (*x*,*y*,*z*), and express all three components of the spin angular momentum vector. The double-cross product in integral (3a) can be transformed to a tensor form: **r** × **v**(**r**) = **r** × (**ω** × **r**) = (**E**$r^2$ − **rr**)·**ω** = **U**(**r**)·**ω**, where **E** stands for the identity tensor, $r = |\mathbf{r}|$ is the length of the positional vector, and **rr** is a dyadic product of the positional vector **r** with itself. The square of the velocity in the Lorentz factor (2b) can be expressed as: $|\mathbf{\beta}(\mathbf{r})|^2 = (\omega^2 r^2 - (\mathbf{\omega}\cdot\mathbf{r})^2)/c^2 = \mathbf{\omega}\cdot\mathbf{U}(\mathbf{r})\cdot\mathbf{\omega}/c^2$. The spin angular momentum then generally reads:

$$\mathbf{S} = \left\{ \int_{R^3} \left(1 - \frac{\mathbf{\omega}\cdot\mathbf{U}(\mathbf{r})\cdot\mathbf{\omega}}{c^2}\right)^{-1/2} \mathbf{U}(\mathbf{r})\rho_m(\mathbf{r})d^3r \right\} \cdot \mathbf{\omega} = \mathbf{I}(\mathbf{\omega})\cdot\mathbf{\omega}, \quad (13)$$

where **I**(**ω**) is the tensor of inertia and **U**(*x*,*y*,*z*) = [[$y^2 + z^2$,-*xy*,-*xz*], [-*xy*,$x^2 + z^2$,-*yz*], [-*xz*,-*yz*,$x^2 + y^2$]] is a symmetric tensor. In contrast with our knowledge about rigid rotators, it is evident that the tensor of inertia is a function of the angular velocity vector **ω**. If we again return to the cylindrical coordinates, having the *z*-axis oriented parallel with the axis of rotation (see Fig. 1), we discover that only diagonal components of the inertia tensor remain and all deviation (off-diagonal) moments vanish. After a lengthy calculation we come to a general result for the diagonal components and their values for the velocity parameter $B_\pm$ = ±1.0821:

$$I_{xx}(B) = I_{yy}(B) = \frac{1}{2}m_0 R^2 \left\{ \frac{3B^2 + 2}{B^3}\mathrm{F}\left(\frac{1}{B}\right) - \frac{1}{B^2} \right\} \Rightarrow I_{xx}(B_\pm) = I_{yy}(B_\pm) = \frac{3}{4}m_0 R^2, \quad (14a)$$

$$I_{zz}(B) = m_0 R^2 \left\{ \frac{B^2 + 2}{B^3}\mathrm{F}\left(\frac{1}{B}\right) - \frac{1}{B^2} \right\} \Rightarrow I_{zz}(B_\pm) = \frac{1}{2}m_0 R^2, \quad (14b)$$

where the imaginary parts have been discarded, F(*x*) is Dawson's function, and subscripts *xx*, *yy*, and *zz* designate rotations around *x*-, *y*-, and *z*-axis, respectively.

Even though the mass density of the rotating rigid body (1) is spherically symmetric, the Poinsot's inertia ellipsoid is only cylindrically symmetric with respect to the *z*-axis. This is a consequence of the fact that the inertia tensor, defined in (13), is a function of the angular velocity. In classical mechanics of rigid rotators, a torque-free precession of an axially symmetric object may occur if the axis of rotation does not coincide with one of the principal axes. In our particular case, the principal axes of the inertia ellipsoid are always parallel or perpendicular to the vector of angular velocity (axis of rotation) which, strictly speaking, rules out any torque-free precession. Nevertheless, it is worthy to discuss the situation of limit precession when the vector of angular velocity and spin angular momentum are deviated from the *z*-axis (symmetry axis) by an infinitesimally small amount.

Let us study a situation depicted in Fig. 3 where the principal axes of the inertia ellipsoid remain parallel with the *x*-, *y*-, and *z*-axis whereas the SAM vector **S** was deflected by some angle *θ* from the *z*-axis. Axes *x*, *y*, and *z* define the body-fixed (internal) reference frame, connected with the spinning top at some time instant. As there is no torque applied to the body, the SAM vector remains constant in space and time, i.e., in the laboratory (external) reference frame. Its direction defines an axis of precession, around which both the *z*-axis and angular velocity vector **ω** orbit counter-clockwise at the precession angular frequency $\Omega_p$. The angular precession velocity vector **Ω**$_p$ can be constructed as a projection of angular velocity **ω** to the direction of SAM vector along the *z*-axis direction (see Fig. 3). When viewed from the body-fixed reference frame, only the *z*-axis remains static being orbited clockwise by the SAM vector and angular velocity at the precession angular frequency $\omega_p$. Evidently, there exists a relation between the angular velocity **ω** and precession angular velocities $\mathbf{\Omega}_p$ and $\mathbf{\omega}_p$ connecting both reference frames: **ω** = $\mathbf{\Omega}_p$ - $\mathbf{\omega}_p$.

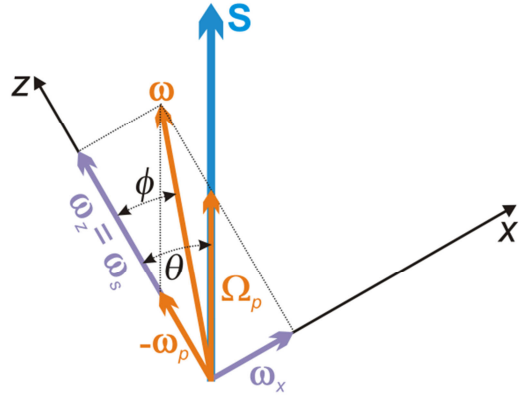

Fig. 3 – A sketch of geometry used in the discussion. Bold variables denote vectors, namely, spin angular momentum **S**, angular velocity **ω**, inverted precession angular velocity -$\mathbf{\omega}_p$ in body-fixed (internal) reference frame, precession angular velocity $\mathbf{\Omega}_p$ in laboratory (external) reference frame, spin angular velocity $\mathbf{\omega}_s$, and projections $\mathbf{\omega}_x$ = **ω** - $\mathbf{\omega}_s$ and $\mathbf{\omega}_z$ = $\mathbf{\omega}_s$ of the angular velocity **ω** to principal axes *x* and *z*. All vectors are in the *xz*-plane and the deflection angles $\theta$, $\phi$ are exaggerated for clarity.

Due to the rotational symmetry, we can restrict ourselves to the *xz*-plane and express the SAM vector as **S** = **I(ω)**·**ω** = $I_{xx}\mathbf{\omega}_x$ + $I_{zz}\mathbf{\omega}_z$ = $I_{xx}\mathbf{\omega}$ + ($I_{zz}$ - $I_{xx}$)$\mathbf{\omega}_s$. Here $I_{xx}$ and $I_{zz}$ are the corresponding diagonal components of the inertia tensor **I(ω)** and $\mathbf{\omega}_x$ = **ω** - $\mathbf{\omega}_s$ and $\mathbf{\omega}_z$ = $\mathbf{\omega}_s$ are projections of the angular velocity to principal axes *x* and *z*. Expressing the spin angular momentum in terms of spin and precession angular velocities, we obtain **S** = $I_{xx}\mathbf{\Omega}_p$ + ($I_{zz}$ - $I_{xx}$)$\mathbf{\omega}_s$ - $I_{xx}\mathbf{\omega}_p$. It follows from the geometry depicted in Fig. 3 that precession angular velocity $\mathbf{\Omega}_p$ is parallel with the SAM vector **S**, whereas $\mathbf{\omega}_p$ is parallel with the spin angular velocity $\mathbf{\omega}_s$ and *z*-axis. This leads to a solution for precession angular velocities: $\mathbf{\Omega}_p$ = **S**/$I_{xx}$ = **ω** + ($I_{zz}/I_{xx}$ - 1)$\mathbf{\omega}_s$ and $\mathbf{\omega}_p$ = ($I_{zz}/I_{xx}$ - 1)$\mathbf{\omega}_s$. Since the spin angular frequency equals $\omega_s = \omega\cos(\phi)$, we can express the two corresponding precession angular frequencies as: $\Omega_p = \omega I_{zz}\cos(\phi)/I_{xx}\cos(\theta)$ and $\omega_p = (I_{zz}/I_{xx} - 1)\omega\cos(\phi)$.

An interesting property of both precession angular frequencies is that they are non-zero for very small and even for zero deflection angles $\theta$ and $\phi$. In the degenerate case of limit precession, when both angles $\theta$ and $\phi$ tend to zero and, in fact, no physical precession occurs, we can neglect both cosines in relationships for precession angular frequencies which gives: $\Omega_p^0 = \omega I_{zz}/I_{xx}$ and $\omega_p^0 = \omega(I_{zz}/I_{xx} - 1)$. Recalling the values (14a) and (14b) for the diagonal components of the inertia tensor and the velocity parameter $B_\pm$, we can express the ratio between the two precession angular frequencies:

$$\left.\begin{array}{l}\Omega_p^0\left(B_\pm\right)=\dfrac{I_{zz}\left(B_\pm\right)}{I_{xx}\left(B_\pm\right)}\omega=\dfrac{2}{3}\omega \\ \omega_p^0\left(B_\pm\right)=\left(\dfrac{I_{zz}\left(B_\pm\right)}{I_{xx}\left(B_\pm\right)}-1\right)\omega=-\dfrac{1}{3}\omega\end{array}\right\}\Rightarrow \dfrac{\Omega_p^0\left(B_\pm\right)}{\omega_p^0\left(B_\pm\right)}=-2\,. \tag{15}$$

Evidently, in the case of limit precession, the relativistic rigid Gaussian body, spinning with the velocity parameter $B_\pm$, precesses twice faster in the external reference frame than in the internal reference frame. This means that after one full revolution (360°) around the external precession axis, defined by the SAM vector, the object does not return to its initial configuration as we would expect. When viewed from the internal reference frame, it will precess only by 180° around the principal $z$-axis. The spinning body will return back into the initial state after two full revolutions (720°) around the external precession axis. In fact, this corresponds to a common perception of spin-1/2 particles.

In summary, bringing all previous findings together, we may claim that the RGB model is capable to explain or predict some important features of the electron, i.e., the rest mass, spin angular momentum, g-factor, magnetic dipole moment, and spin-1/2 appearance. This in other words means, we have arrived at Dirac's picture of an electron but in classical terms.

*Note regarding the electron radius and angular frequency:* In previous calculations the 1/e radius $R$, determining the width of the mass and charge density, was not needed. In fact, it can be derived from Eq. (6a) since we know that the spin angular momentum of an electron is $S_z = \pm\hbar/2 = m_0cRB_\pm/2$. From here and from the definition of the velocity parameter, we obtain values for both the radius $R$ = 356.86 fm and angular frequency $\omega$ = 909.03 rad/as. The radius $R$ is in close relation to the so called reduced Compton wavelength of an electron $\lambdabar_C$ = 386.16 fm since it holds true that $R = \lambdabar_C/|B_\pm|$.

## 5. The static electromagnetic field

Until now we have been discussing mainly the mechanical properties of a rigidly spinning Gaussian body with inclusion of relativistic effects. The analysis of the magnetic dipole moment and g-factor has confirmed that the mass and charge density are of equal Gaussian shape and 1/e radius. Since the charge density is in steady rotation, a static electromagnetic (EM) field is created. In order to derive the spatial shape of the EM field, we have to start with general equations for Liénard-Wiechert potentials (c.f. [28]):

$$\Phi(\mathbf{r},t)=\dfrac{1}{4\pi\varepsilon_0}\iiint_{R^3}\int_R\dfrac{\rho_e(\mathbf{r}',t')}{|\mathbf{r}-\mathbf{r}'|}\delta\left(t-t'-\dfrac{|\mathbf{r}-\mathbf{r}'|}{c}\right)dt'd^3r'\,, \tag{16a}$$

$$\mathbf{A}(\mathbf{r},t)=\dfrac{\mu_0}{4\pi}\iiint_{R^3}\int_R\dfrac{\mathbf{j}_e(\mathbf{r}',t')}{|\mathbf{r}-\mathbf{r}'|}\delta\left(t-t'-\dfrac{|\mathbf{r}-\mathbf{r}'|}{c}\right)dt'd^3r'\,, \tag{16b}$$

where $\Phi(\mathbf{r},t)$ and $\mathbf{A}(\mathbf{r},t)$ is the scalar and vector potential at observer's position $\mathbf{r}$ and time $t$, $\varepsilon_0 = 8.8542\cdot10^{-12}$ F/m and $\mu_0 = 1.2566\cdot10^{-6}$ H/m is the permittivity and permeability of vacuum, and $\rho_e(\mathbf{r}',t')$ and $\mathbf{j}_e(\mathbf{r}',t')$ is a general charge and current density at some position $\mathbf{r}'$ and retarded time $t'$. The potentials satisfy the Lorentz gauge condition defined as: div $\mathbf{A}(\mathbf{r},t) + c^{-2}\partial_t\Phi(\mathbf{r},t) = 0$. The condition of steady rotation greatly simplifies the solution of these integrals since the charge and current density, as defined in Eqs. (10a) and (10b), are no longer functions of time. Therefore, the delta-function in integrals (16a) and (16b) vanishes when integrated over the retarded time $t'$. Inserting from Eqs. (10a) and (10b), we find the Liénard-Wiechert potentials of the extended Gaussian electron in analytical forms:

$$\Phi(\mathbf{r}) = \frac{q}{4\pi\varepsilon_0 r}\operatorname{erf}\left(\frac{r}{R}\right), \tag{17a}$$

$$\mathbf{A}(\mathbf{r}) = \frac{q\mu_0}{8\pi}\left\{\frac{R^2}{r^2}\operatorname{erf}\left(\frac{r}{R}\right) - \frac{2}{\sqrt{\pi}}\frac{R}{r}\exp\left(-\frac{r^2}{R^2}\right)\right\}\boldsymbol{\omega}\times\mathbf{r}_0, \tag{17b}$$

where $r = |\mathbf{r}|$ and $\mathbf{r}_0 = \mathbf{r}/r$ is the magnitude and direction of the positional vector, respectively. Evidently, at the limit of small 1/e radius ($R \to 0$), i.e., at the limit of a point-like particle, the scalar potential (17a) tends to the Coulomb potential $\Phi(\mathbf{r}) \to q/4\pi\varepsilon_0 r$ and the vector potential to zero $\mathbf{A}(\mathbf{r}) \to 0$. Furthermore, the vector potential will be zero for zero angular velocity $\boldsymbol{\omega} = 0$ whereas the scalar potential is independent of rotation. A very important feature of these potentials is an absence of infinities at the origin and asymptotic Coulomb-like behavior of the scalar potential at large distances. Calculating the limits of these potentials at the origin, we obtain: $\lim_{r\to 0}\Phi(\mathbf{r}) = q/2\pi^{3/2}\varepsilon_0 R$ and $\lim_{r\to 0}\mathbf{A}(\mathbf{r}) = 0$. This ensures that the corresponding electric and magnetic fields are finite and the related electromagnetic energy density is integrable. The difference between the Coulomb potential and the scalar potential defined by (17a) becomes significant at distances comparable or shorter than the radius R, i.e., reduced Compton wavelength. At the radial distance of 2R, the relative difference falls below 0.5%; for the radial distance equal to Bohr radius $a_0 = 52.9$ pm, the relative difference is approximately equal to 1 - erf(148) and, therefore, negligibly small.

In conditions of the Lorentz gauge, the general electric $\mathbf{E}(\mathbf{r},t)$ and magnetic $\mathbf{B}(\mathbf{r},t)$ field can be expressed by differentiating the scalar and vector potential as follows: $\mathbf{E}(\mathbf{r},t) = -\operatorname{grad}\Phi(\mathbf{r},t) - \partial_t \mathbf{A}(\mathbf{r},t)$ and $\mathbf{B}(\mathbf{r},t) = \operatorname{rot}\mathbf{A}(\mathbf{r},t)$. Since the potentials are not functions of time, these two equations will reduce to: $\mathbf{E}(\mathbf{r}) = -\operatorname{grad}\Phi(\mathbf{r})$ and $\mathbf{B}(\mathbf{r}) = \operatorname{rot}\mathbf{A}(\mathbf{r})$, respectively. From here we can derive the EM-field in the following forms:

$$\mathbf{E}(\mathbf{r}) = \frac{q}{4\pi\varepsilon_0 R^2}\left\{\frac{R^2}{r^2}\operatorname{erf}\left(\frac{r}{R}\right) - \frac{2}{\sqrt{\pi}}\frac{R}{r}\exp\left(-\frac{r^2}{R^2}\right)\right\}\mathbf{r}_0 = E(\mathbf{r})\mathbf{r}_0, \tag{18a}$$

$$\mathbf{B}(\mathbf{r}) = \frac{R^2}{2c^2}\left\{\left(\frac{\rho_e(\mathbf{r})}{\varepsilon_0} - \frac{E(\mathbf{r})}{r}\right)\boldsymbol{\omega} + \left(\frac{3E(\mathbf{r})}{r} - \frac{\rho_e(\mathbf{r})}{\varepsilon_0}\right)(\boldsymbol{\omega}\cdot\mathbf{r}_0)\mathbf{r}_0\right\}, \tag{18b}$$

where $E(\mathbf{r}) = \mathbf{r}_0 \cdot \mathbf{E}(\mathbf{r})$ is the magnitude of the electric field at position $\mathbf{r}$. Evidently, the electric field is spherically symmetric whereas the magnetic field is only cylindrically symmetric with respect to the direction of the angular velocity vector $\boldsymbol{\omega}$. Cross-sections through the electric and magnetic field are depicted in Fig. 4. As shown in Figs. 4(a) and 4(b), the absolute magnitude of the electric field $|\mathbf{E}(\mathbf{r})|$ exhibits a local zero minimum at the origin and global maximum $4.8 \cdot 10^{15}$ V/m located 345.4 fm off the center. As depicted in Figs. 4(c) and 4(d), the absolute magnitude of the magnetic field $|\mathbf{B}(\mathbf{r})|$ has a local zero minimum in a ring-like volume, located 539.3 fm off the z-axis (axis of rotation), whereas a value as high as 30.7 MT is reached in the global maximum at the origin. It can be easily proven that Eqs. (18a) and (18b) comply with time-independent Maxwell equations, i.e., $\operatorname{rot}\mathbf{E}(\mathbf{r}) = 0$, rot $\mathbf{B}(\mathbf{r}) = \mu_0 \mathbf{j}_e(\mathbf{r})$, div $\mathbf{E}(\mathbf{r}) = \rho_e(\mathbf{r})/\varepsilon_0$, and div $\mathbf{B}(\mathbf{r}) = 0$. Finally, by comparing Eqs. (17b) and (18a), we find an interesting relation between the vector potential and electric field $\mathbf{A}(\mathbf{r}) = (R^2/2c^2)\, \boldsymbol{\omega} \times \mathbf{E}(\mathbf{r})$.

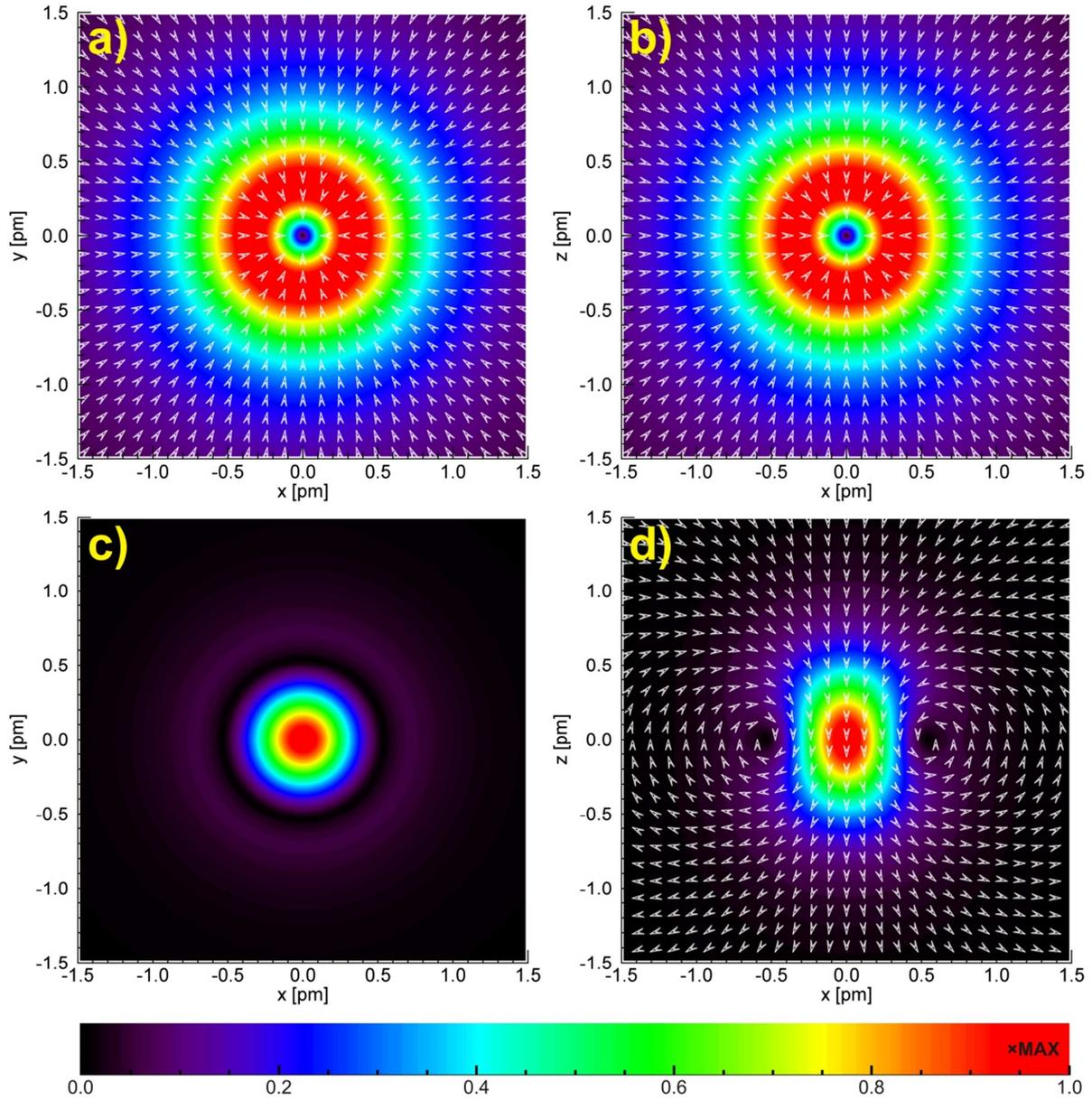

Fig. 4 - Cross-sectional views through electric and magnetic field in *xy* and *xz* planes depicting both the absolute magnitude (as a colored contour plot) and directions of field lines (as arrows). Cross-sections through the electric field in *xy*-plane (a) and *xz*-plane (b) confirm its spherical symmetry whereas cross-sections through the magnetic field in *xy*-plane (c) and *xz*-plane (d) show the cylindrical symmetry. Cross-sections in *yz*-plane are identical to those from *xz*-plane in both cases. The vector of angular velocity points in the *z*-direction. No arrows in figure (c) indicate that the magnetic field is perpendicular to the *xy*-plane albeit its direction changes from antiparallel to parallel with the *z*-axis when crossing the local off-axis minimum. Each color in the color scale corresponds to a value at a certain level of maximum, i.e., of $4.8 \cdot 10^{15}$ V/m and 30.7 MT for electric and magnetic field, respectively.

## 6. Electromagnetic energy and Poynting vector

It follows from the above analysis that the spinning Gaussian electron possesses both the electric and magnetic field. Hence there must be a certain, presumably finite, portion of energy attributable to these fields. Moreover, one can define the Poynting vector at every point of space, which is associated with the energy flow. Recalling definitions of the electromagnetic energy density $u(\mathbf{r})$ and Poynting vector $\mathbf{P}(\mathbf{r})$ [28]:

$$u(\mathbf{r}) = u_E(\mathbf{r}) + u_B(\mathbf{r}) = \frac{\varepsilon_0}{2}|\mathbf{E}(\mathbf{r})|^2 + \frac{1}{2\mu_0}|\mathbf{B}(\mathbf{r})|^2, \tag{19a}$$

$$\mathbf{P}(\mathbf{r}) = \frac{1}{\mu_0}\mathbf{E}(\mathbf{r}) \times \mathbf{B}(\mathbf{r}), \tag{19b}$$

inserting from Eqs. (10a), (18a), and (18b), and integrating the electric $u_E(\mathbf{r})$ and magnetic $u_B(\mathbf{r})$ energy density over the entire space, we obtain:

$$U = U_E + U_B = \frac{\alpha \hbar c}{\sqrt{2\pi}R} + \frac{\alpha \hbar c B^2}{6\sqrt{2\pi}R}, \tag{20a}$$

$$\mathbf{P}(\mathbf{r}) = \frac{\varepsilon_0 R^2}{2}\left(\frac{E(\mathbf{r})}{r} - \frac{\rho_e(\mathbf{r})}{\varepsilon_0}\right)\boldsymbol{\omega} \times \mathbf{E}(\mathbf{r}). \tag{20b}$$

Here $B = \omega R/c = kR$ is the velocity parameter and $\alpha = q^2/4\pi\varepsilon_0\hbar c = 7.2974 \cdot 10^{-3}$ is the fine-structure constant. Evidently, for non-zero 1/e radius $R$ the total electromagnetic energy $U$, as well as its components $U_E = \alpha\hbar c/(2\pi)^{1/2}R$ and $U_B = U_E B^2/6$, is finite. For the radius $R = 356.86$ fm and angular frequency $\omega = 909.03$ rad/as, i.e., for the velocity parameter $B_\pm = \pm 1.0821$, we obtain the constituent energies associated with the electric field $U_E = 1609.7$ eV and the magnetic field $U_B = 314.15$ eV. Hence the total electromagnetic energy $U = 1923.9$ eV of the Gaussian electron represents only 0.38% of its rest energy. It should be also noted that in the case of a point-like particle, i.e., for $R \to 0$, the energy of the electric field diverges whereas the energy of the magnetic field tends to zero.

Since the electromagnetic energy as well as the Poynting vector is non-zero, it should be discussed whether or not the particle radiates when in steady rotation. Let us begin with the general Poynting's theorem putting into relation the electromagnetic energy density, Poynting vector, and absorbed power: $\partial_t u(\mathbf{r},t) + \text{div } \mathbf{P}(\mathbf{r},t) = -\mathbf{j}_e(\mathbf{r},t)\cdot\mathbf{E}(\mathbf{r},t)$. The first term, as well as all other time dependencies, in this equation vanishes since the rotational motion is steady and thus time-independent. Furthermore, it follows from Eqs. (10b) and (18a) that the term on the right-hand side is zero as the current density is perpendicular to the electric field everywhere in the space. It is straightforward to show that divergence of the Poynting vector (the second term) is identically equal to zero and, therefore, the Poynting theorem is proven to be valid. Since all the terms are zero, there cannot be any radiation emerging from the spinning body under question unless it is disturbed from the steady state of motion.

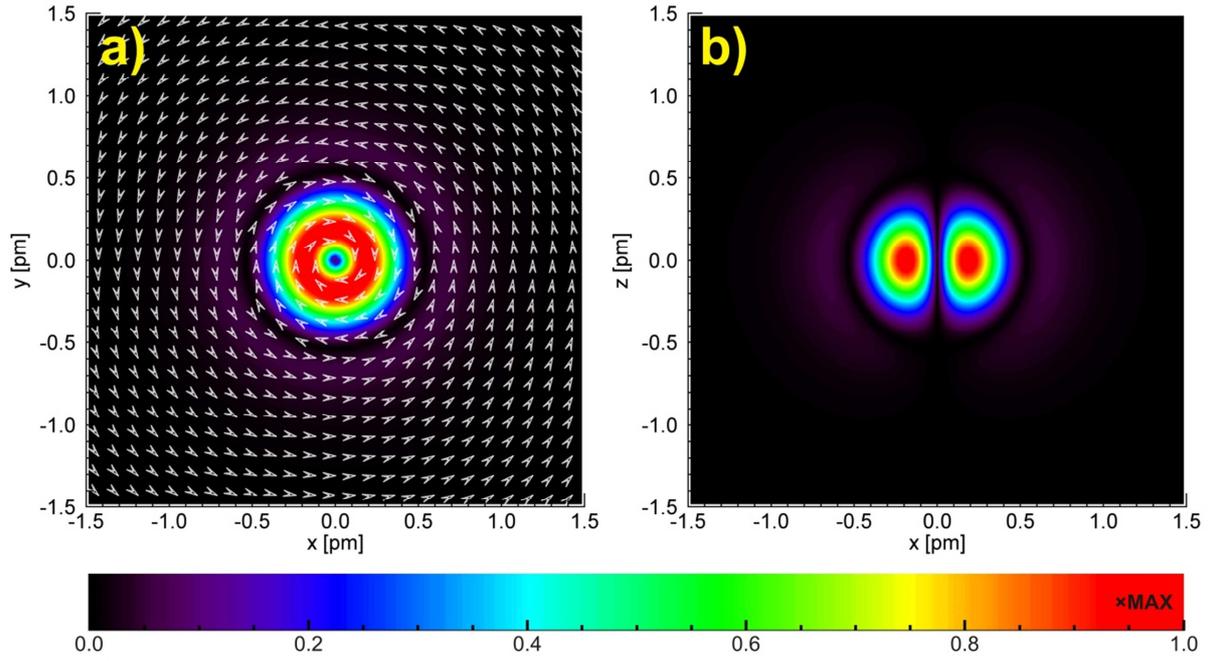

Fig. 5 - Cross-sectional views of the Poynting vector field in *xy* (a) and *xz* (b) plane depicting both the absolute magnitude (as a colored contour plot) and directions (as arrows) of local Poynting vectors. The cross-section in the *yz*-plane is identical to those from *xz*-plane and, therefore, not displayed. The vector of angular velocity points in the *z*-direction. No arrows in figure (b) indicate that all vectors are perpendicular to the *xz*-plane. Each color in the color scale corresponds to a value at a certain level of maximum.

## 7. Electromagnetic correction to the g-factor ($g_e^*$ = 2.0021)

The Poynting vector field, as defined in Eqs. (19b) and (20b) and depicted in Fig. 5, can be associated with local momentum density of the electromagnetic field through an equation: $d\mathbf{p}_{EM}(\mathbf{r})/dV = \mathbf{P}(\mathbf{r})/c^2$. In accord with the first definitional integral in Eq. (3a), an angular momentum can be ascribed to such a momentum density field by solving:

$$\mathbf{S}^{EM} = \varepsilon_0 \mu_0 \int_{R^3} \mathbf{r} \times \mathbf{P}(\mathbf{r}) d^3 r . \tag{21}$$

For the vector of angular velocity being parallel with the *z*-axis, only the *z*-component of the electromagnetic angular momentum $S_z^{EM}$ remains non-zero. By inserting from Eq. (20b) to (21) and integrating over the entire space, we obtain:

$$S_z^{EM}(B) = -\frac{\alpha \hbar B}{3\sqrt{2\pi}} . \tag{22}$$

Evidently, the electromagnetic contribution to the total angular momentum is very small but negative since the vector of the electromagnetic angular momentum $\mathbf{S}^{EM}$ points against both the spin angular momentum $\mathbf{S}$ and its non-relativistic limit $\mathbf{S}^{nR}$.

In order to incorporate the electromagnetic correction into the g-factor, we have to remember how the g-factor $g_e$ = 2 was derived. In fact, the g-factor represents a multiplication parameter which inserts a relation between non-relativistic and relativistic spin angular momenta. For the velocity parameter $B_\pm = \pm 1.0821$, the non-relativistic spin angular momentum is exactly twice that of relativistic $S_z^{nR}(B_\pm) = 2S_z(B_\pm)$, whereby the g-factor $g_e = 2$ appears in the equation for the magnetic dipole moment. In this particular case, the spinning body corresponds to Dirac's ideal picture of an electron. However, by introducing the electromagnetic correction, we inevitably distort this picture and change some parameters or properties of the particle. Detailed analysis of this problem is a very difficult task which requires a deep grasp of fundamental principles regarding how the electromagnetic energy couples into the total, experimentally measurable, rest energy (mass). This will most likely lead to the concept of the so called electromagnetic and mechanical mass, discussed, for example, by Schwinger [30] and MacGregor [16]. Nevertheless, in the analysis presented here we prefer to avoid the concept of EM mass since it might be too counterintuitive. Instead, we will attempt to work with quantities which were already derived.

It follows from the existence of the electromagnetic angular momentum that the mechanical spin angular momentum $S_z(B)$ can no longer keep its value, given by Eqs. (4a) and (6a), and must be replaced by a modified spin angular momentum $S_z^*(B)$. Since it holds true that the total spin angular momentum $\hat{S}_z(B)$ of the particle under question must always be equal to $\pm\hbar/2$, we can write $\hat{S}_z(B) = S_z^*(B) + S_z^{EM}(B) = \pm\hbar/2$. Any modification to relativistic $S_z(B)$ must inevitably imply a modification to its non-relativistic limit $S_z^{nR}(B)$. Therefore, it is legitimate to write $\hat{S}_z^{nR}(B) = S_z^{nR*}(B) + S_z^{EM}(B) = \pm\hbar$, where $\hat{S}_z^{nR}(B)$ and $S_z^{nR*}(B)$ is the total and modified non-relativistic SAM, respectively. The electromagnetic angular momentum $S_z^{EM}(B)$, as defined in Eq. (22), is generally valid in both the non-relativistic and relativistic regime of rotation; hence it adds to both modified momenta by the same amount. By analogy with Eq. (11), the modified non-relativistic SAM $S_z^{nR*}(B) = \pm\hbar - S_z^{EM}(B)$ is related to the magnetic dipole moment and the g-factor is measured in units of the total spin angular momentum $\hat{S}_z(B) = \pm\hbar/2$. Therefore, provided that the velocity parameter remains unchanged ($B = B_\pm = \pm 1.0821$), the modified g-factor can be expressed as:

$$g_e^* = \frac{S_z^{nR*}(B_\pm)}{\hat{S}_z(B_\pm)} = \frac{\pm\hbar - S_z^{EM}(B_\pm)}{\pm\hbar/2} = 2.002100133. \tag{23}$$

Albeit the value of the corrected g-factor is very close to the measured one ($g_e = 2.0023193043617$), it is still far from the QED accuracy since only the electromagnetic correction has been considered. Higher-order corrections will probably emerge from interaction of the particle with the surrounding non-empty environment, usually denoted as QED vacuum. Furthermore, it should be noted that the phenomenon of the electromagnetic mass was not fully understood and was rather bypassed by a much simpler analysis. This could also limit the accuracy of this calculation. Nonetheless, taking into account that all calculations were performed in a purely classical manner, we may consider this result as very good.

## 8. Conclusions

It has been shown that the g-factor of an electron can be, to a very high accuracy, explained classically without involvement of quantum mechanics or quantum electrodynamics. The model of a relativistic spinning rigid Gaussian body, presented in this paper, has shown the capability to reproduce or explain many parameters and properties of Dirac's electron including its rest mass, spin angular momentum, magnetic dipole moment, g-factor ($g_e = 2$), spin-1/2 behavior, and intrinsic electromagnetic field. By introducing the electromagnetic correction to the mechanical spin angular momentum, a corrected value of the g-factor ($g_e^*$ = 2.0021) was obtained. The detailed analysis of the presented model supports the conclusion of Cohen [13] that the g-factor greater than one can be explained in terms of special relativity applied to systems of equal mass and charge density.

The intention of this paper is by no means to compete with or contradict excellent results of quantum electrodynamics but rather to show an alternative and instructive way which could potentially help to gain a deeper insight into the physics of electron spin and connected underlying phenomena.

**Acknowledgments**

The author thanks the Academy of Sciences of the Czech Republic for postdoctoral financial support.